\documentclass[fleqn,10pt]{wlscirep}
\usepackage[utf8]{inputenc}
\usepackage[T1]{fontenc}

\usepackage{graphicx}
\usepackage{booktabs}

\usepackage{indentfirst}
\title{Crash dynamics of interdependent networks}

\author[1,2]{Jie Li}
\author[1,2,*]{Chengyi Xia}
\author[3,4,*]{Gaoxi Xiao}
\author[5,6,7]{Yamir Moreno}
\affil[1]{Tianjin Key Laboratory of Intelligence Computing and Novel Software Technology, Tianjin University of Technology, Tianjin 300384, China}
\affil[2]{Key Laboratory of Computer Vision and System (Ministry of Education), Tianjin University of Technology, Tianjin 300384, China}
\affil[3]{School of Electrical and Electronic Engineering, Nanyang Technological University, Singapore 639798}
\affil[4]{Complexity Institute, Nanyang Technological University, Singapore 639798}
\affil[5]{Instituto de Biocomputaci\'on y F\'isica de Sistemas Complejos (BIFI), Universidad de Zaragoza, Zaragoza 50018, Spain}
\affil[6]{Departamento de F\'isica Te\'orica, Facultad de Ciencias, Universidad de Zaragoza, Zaragoza 50009, Spain}
\affil[7]{ISI Foundation, Turin, Italy}

\affil[*]{xialooking@163.com, EGXXiao@ntu.edu.sg}


\begin{abstract}
The emergence and evolution of real-world systems have been extensively studied in the last few years. However, equally important phenomena are related to the dynamics of systems' collapse, which has been less explored, especially when they can be cast into interdependent systems. In this paper, we develop a dynamical model that allows scrutinizing the collapse of systems composed of two interdependent networks. Specifically, we explore the dynamics of the system's collapse under two scenarios: in the first one, the condition for failure should be satisfied for the focal node as well as for its corresponding node in the other network; while in the second one, it is enough that failure of one of the nodes occurs in either of the two networks. We report extensive numerical simulations of the dynamics performed in different setups of interdependent networks, and analyze how the system behavior depends on the previous scenarios as well as on the topology of the interdependent system. Our results can provide valuable insights into the crashing dynamics and evolutionary properties of interdependent complex systems.
\end{abstract}

\begin{document}

\flushbottom
\maketitle
%
%
%

\section*{Introduction}
\label{sec:0}

Many complex systems, both in nature and human society, undergo a cycle that starts at birth and finish with their eventual extinction or collapse after a period of stabilization\cite{song2006natphys,Castellano2009}. Over the past two decades, advances in areas such as network science\cite{Wangxf2003,boccaletti2006report}, have led to many researches which have placed the focus mainly on the first part of this cycle, namely, on characterizing the emergence and growth of complex networked systems. In particular, the availability of new data made it possible a more realistic characterization of the structure of complex systems \cite{watts98nature,barabasi99science}, which in turn triggered many other studies \cite{Albert2002,boccaletti2014report,ISI:000407664300011,gaozk2017,yamir2013multilayer,yamir_review1} that have uncovered additional hidden topological features, as well as the evolution and dynamical properties of such complex systems. This is the case of evolutionary game models widely used to study cooperation\cite{ISI:000274868700004,ISI:000417373200033,ISI:000356901900015,meloni2017,review07,review09,review10}: the incorporation of population structure in the form of networks have led to new insights and significant progresses in our understanding of how cooperative behavior emerges and evolves in human, animal, and even machine populations. Other examples worth mentioning are given by the remarkable increase in our ability to describe disease spreading \cite{ISI:000360273200001,ISI:000390637400001,yamir_review2}, in detecting community structure  \cite{community2009review,newman2006pnas,ISI:000384234700007} and {\color{black}even in depicting the information diffusion in social networks \cite{add2_pnas}}, to mention a couple of relevant examples. On the other hand, cascading patterns in complex systems are ubiquitous and usually lead to function loss in these systems. Examples of processes in which cascades play a key role include  {\color{black}the \emph{k}-core organization of randomly damaged complex networks \cite{add1_prl}}, power-grid failures \cite{ISI:000247625600036}, flash crashes in financial markets \cite{stafford2013flash} and the spread of political movements \cite{d2017curtailing} such as "Arab Spring", etc. Moreover, many real systems are often interconnected or interdependent, e.g., the mutual underpinning between power-grids and communication systems, which further aggravates the risk of cascading failure taking place upon them.

The analysis of cascading failure of interrelated systems has also attracted quite a lot of interest in network science \cite{Brummitt2015}. This line of research was preceded by works \cite{ISI:000088383800038,ISI:000165399000054,ISI:000165884200052} in which cascading processes that occurred on top of single networks were studied, reporting effects such as that scale-free (SF) networks exhibit robust-yet-fragile properties, that is, SF networks are robust against random failures but fragile when suffering targeted attacks. More recently, it has also been shown that for the same kind of cascading dynamics, interdependency among multiple networked systems may surprisingly weaken the robustness against random attacks. In fact, the failure of a node within one system may lead to function loss of both its nearest neighbors within its network and its connections to the other coupled system \cite{ISI:000276635000035,ISI:000369085100035}. Yet, our understanding of cascading failures in complex systems has been driven mostly by theoretical studies, including the seminal Bak-Tang-Wiesenfeld sandpile model of self-organized criticality \cite{Bak1987}. Despite the existence of a few realistic experiments, the lack of data has made it hard to compare on quantitative grounds, real-world cascading failures and model predictions. A very recent work \cite{Yang2017}, however, has opened the path to bridge this gap, where Yang et al\cite{Yang2017} used data from North-American power grids to show that there exists a vulnerable set of nodes in the system that mainly consists of a small but topologically central portion of the network, with large cascades disproportionately emerging at or triggered by initial failures close to this set.

Despite these recent advances, a current challenge is to develop a general approach to model the dynamics of cascading failures. To this end, Yu et al.\cite{ISI:000385610400047} proposed a simple network-based system-collapsing model aimed at characterizing the evolution of complex systems and their collapse, {\color{black}which was termed as $KQ$-cascade model. Here, any node will leave the system with a fixed probability once its degree is less than a specified value $k_s$ (\emph{i.e.},$K$-condition) or the proportion of lost nearest neighbors is higher than $q$ (that is, $Q$-condition)}. Interestingly enough, the model is suited to describe pseudo-steady state during the system's collapse, assuming that any individual node will adopt a specific action just based on local information about the vicinity of it. {\color{black}Among them, as defied in Ref. \cite{ISI:000385610400047}, the pseudo-steady state represents the situation that the networks seemingly appear to be rather stable and only lose a few nodes at each time step; after a long period, however, the systems suddenly crash, sometimes within a few steps. Accordingly,} this model cannot only explain the sudden collapse of the system, but also it can account for meta-stable phenomena during the course of system's degradation. However, the previous model assumes that nodes leave the network with the same probability once the KQ condition is satisfied regardless of their degrees. Recently, an improved version of the $KQ$-cascade model \cite{modifiedKQ2018} was studied, in which it is assumed that this probability is correlated with the nodes' degrees. Apart from adding more realism to the model since node heterogeneity is a defining feature of many real-world networks, numerical simulations showed\cite{modifiedKQ2018} that the crashing behavior can be greatly affected by this kind of heterogeneity.

Here, we take one step forward and add even more realism to the previous cascading model by studying how the model behaves on top of interdependent systems\cite{ISI:000276635000035,ISI:000369085100035,boccaletti2014report} such as power grids and communication systems, in which power grids provide the power supply for communication ones while communication networks control and transfer data to the power grid network. We study the system behavior during the process of collapse, including the characterization of how metastable phenomena change and whether sudden collapses of the system are possible. Our results report numerical simulations of the generalized model\cite{modifiedKQ2018} for interdependent systems made up of two layers or networks that can have either homogeneous or heterogeneous topologies. These simulations show that the rule governing how nodes detach from the networks determines the system crashing behavior and its robustness.

\section*{Methods and Models}
\label{sec:3}

The dynamics of collapse is determined by two main components in our model: the first one defines the cascading process within one single network and the second one accounts for coupling patterns between the two interdependent networks. For the cascading process, we follow the model proposed in Ref.\cite{ISI:000385610400047} on top of a graph $G=(V, E)$, where $V$ denotes the set of nodes and $E$ represents the set of links. Additionally, for the second ingredient, we consider different kinds of topologies as well as that the two networks that made up the whole interdependent system are linked following different degree correlations between the nodes of the two layers.

Specifically, we first generate three classes of random networks of size $N=10^4$ by using the configuration model: Erd\"os and R\'enyi (ER) networks with an average degree $<k>=20$, exponential (EXP) networks with an average degree $<k>=20$ and a degree cutoff of $100$, and scale-free networks with a power exponent $\gamma=2$, a minimum degree of $3$ and maximum degree of $100$.
We also set for each network the parameters that will determine whether a node fail or not, namely, the the critical degree $k_s$ and the fraction of link loss $q$. Then, {\color{black}similar to the consideration of inter-layer degree correlation in previous works \cite{add3_pre,add4_sr,add5_pre}, we construct two-layered correlating networks by the following procedure: firstly} rank the nodes within each network according to their degrees, and then build the coupling relationship between nodes on two different networks by adopting one of three possible scenarios:
\begin{itemize}
\item \textbf{One-to-one mapping with assortative degree sequence:} The nodes in one network are positively correlated with the nodes in the other network according to their degrees, that is, the node with the maximum degree on one network is related to the node with the maximum degree on the other layer.
\item \textbf{One-to-one mapping with disassortative degree sequence:} The nodes in one network are negatively correlated with the nodes in the other network according to their degrees, that is, the node with the maximum degree on one network is related to the node with the minimum degree on the other layer.
\item \textbf{One-to-one mapping with random degree sequence:} each node in one layer is linked randomly with any other node of the other network.
\end{itemize}

{\color{black}Notably, this mapping relationship is unique, that is, interlayer links are one-to-one. However, the actual interlayer links between corresponding node pairs on the two layers do not exist, here we just build one kind of virtual mapping patterns as mentioned above.}

As for the dynamics, for each network and for each step of the simulation, any node fulfilling either the condition $k<k_s$ or that the fraction of links lost is greater than $q$ might be switched off (or leave) the network it belongs to as specified below. Since here we deal with an interdependent system, we also have to set a rule for linking the cascading failure in one layer with the dynamics of the other layer. In other words, we need to define how the two layers are dynamically interdependent. Here, we propose two scenarios of node detachment during the cascading process:
\begin{itemize}
\item \textbf{"AND" scenario:} A pair of nodes choose to leave the system with the given probability $f$ if and only if both of them simultaneously meet the $KQ$-cascade condition within their respective layer -or network.
\item \textbf{"OR" logic:} A pair of nodes will be removed from the current system with a fixed probability $f$ if any of them fulfills the $KQ$-cascade condition within its corresponding network.
\end{itemize}

In addition, we explore the impact of network setups on the breakdown of interdependent networks. To this end, we consider two different setups regarding the topology of the interdependent networks: one scheme assumes that the class of networks that made up the interdependent system is the same, while the network classes are different in the other scheme. For the sake of presentation, we call these two schemes as "$multiType=1$" and "$multiType=2$", respectively, as it is summarized in Table \ref{tab:network}.

\begin{table}[ht]
\centering
\begin{tabular}{p{100pt}p{60pt}}
\toprule
$multitype=1$ & $multitype=2$ \\
\midrule
$ER - ER$ & $EXP - ER$ \\
$EXP - EXP$ & $SF - EXP$ \\
$SF - SF$ & $SF - ER$ \\
\bottomrule
\end{tabular}
\caption{\label{tab:network} Network classes that make up the two-layer interdependent networks studied in this paper. If two networks are of the same class, we refer to them as $multiType=1$, otherwise, this scheme is called $multiType=2$.}
\end{table}
{
\color{black}
Accordingly, the complete cascading process within two interdependent networks can be summarized as follows
\begin{itemize}
  \item Initialization: generate the expected random network topology by the configuration model; record the original degree $k_i$ for each node; set the model parameter $k_s$ and $q$; build the coupling relationship between two interdependent networks according to the simulation requirements; determine the detachment logic ("AND"/"OR");
  \item The first simulation step: for any pair of nodes on the two networks, judge whether this pair of nodes satisfy the condition: if $k_i<k_s$ within respective networks, this pair of nodes shall be removed from the networks with the given probability $f$; otherwise, this node pair shall stay in the network;
  \item Iterations in the following steps: count the current degrees for all nodes within two-layered networks as recorded in the initialization. Once a pair of nodes have their current degrees lower than $k_s$ or have lost more than $q$ fraction of original neighbors within at least one of the two layers, we will further check the detachment condition in accordance with the given logic and then determine whether this pair of nodes will leave the network simultaneously with a fixed probability $f$. After all node pairs have been checked, the degree of all nodes will be updated at the end of this time step.
\end{itemize}

The above procedure shall be repeated until no further nodes can be removed from the two-layer network.
}

\section*{Results}
\label{sec:1}

We have performed extensive numerical simulations of the dynamics described in the previous section on top of interdependent networks generated as described earlier. In this section, we present the results of exploring the system's dynamics using the generalized $KQ$-cascade model, and report our findings for several network configurations and the two scenarios for the dynamical failure of the nodes explained before (i.e., "AND" and "OR" schemes). {\color{black}Meanwhile, the robustness of system can be measured as the fraction of remaining node pairs in the network at the stationary state; that is, the higher this ratio is, the more robust the system is.}

\subsection*{Fraction of remaining node pairs in the network.}
\begin{figure}[htb]
\centering
\includegraphics[width=1.0\columnwidth]{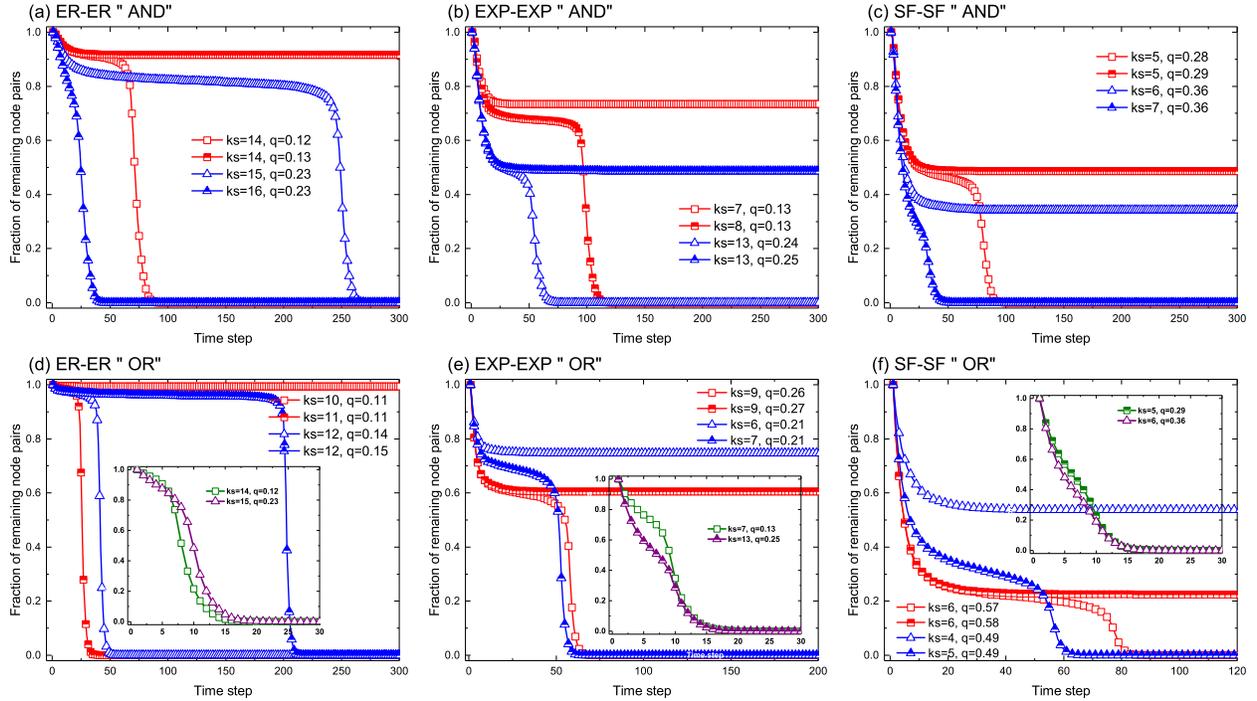}
\caption{Fraction of node pairs that remains in the whole network as a function of time for different model parameters and several configurations of the interdependent system as indicated. On the top three panels, the detachment rule for an interlayer pair to leave the network is the "AND" logic, while the "OR" logic was adopted for the three panels of the bottom. The probability of removing nodes one they meet the detachment rule has been set to $f=0.2$. The network parameters are:  $N=10^4$, $<k>=20$ for ER random graphs, the exponential networks have an average degree of $<k>=20$ and a degree cutoff of $100$; and scale-free networks are generated with $\gamma=2$, a minimum degree of $3$ and a degree cutoff of $100$.}
\label{fig:multitype_1}
\end{figure}

First, we illustrate the fraction of nodal pairs that remains within the interdependent networks at each time step for a fixed value of $k_s$ and $q$ in Fig.\ref{fig:multitype_1}. The two-layered networks are of the same class but with different topologies (i.e., $multitype=1$) and the interlayer degree correlations correspond to the assortative scenario [results for the other two cases are provided in the Supplementary Information(SI)]. In the top three panels [(a),(b),(c)], we show results for the "AND" rule, which is applied at each iterative time step, that is, a pair of nodes will be removed from the network with probability $f$ if and only if both of the corresponding partners on two interdependent networks fulfill the leaving condition. Conversely, the results shown in the three panels at the bottom of the figure [(d),(e),(f)] were obtained by using the "OR" rule. These results indicate that the cascading process in the interdependent system exhibits the similar behavior as that observed in single layer networks, in which a slight change in $k_s$ or $q$ around their critical values leads to an abrupt collapse of the network. In other words, here we also observe the existence of a meta-steady state and a sudden crash for the case of interdependent two-layer networks. For instance, this is the behavior observed when the loss-tolerance parameter $q$ decreases from $0.13$ to $0.12$ keeping $k_s=14$ (red curves in panel (a)): the system goes from a situation in which at the stationary state roughly 90\% of the nodes are intact to the full collapse at $q=0.12$. Similar phenomena can also be observed in panels (b) or (c) for exponential and SF networks, albeit for other values of the parameters. A comparison between the $multitype=1$ configurations shows that the $ER-ER$ setup is more robust when compared to $EXP-EXP$, which in its turn is more robust than the $SF-SF$ configuration $-$ the most fragile one of the three settings.


On the other hand, by comparing panels (a), (b) and (c) with, respectively, the insets of panels (d), (e) and (f) $-$which have been obtained utilizing the same parameters to make the direct comparison possible$-$, it can be seen that, as far as the fraction of remaining interlayer pairs is concerned, the "AND" rule renders the system more resilient than the "OR" scenario. This is somehow expected given the more constrained "AND" scenario that needs the cascading conditions to be satisfied in both networks. Moreover, it is also possible to find for the "OR" scenario regions of parameters in which the dynamical behavior obtained for the "AND" rule is also reproduced. Namely, as shown in panels (d), (e) and (f) there are scenarios in which the system collapses in a few time steps, others in which there is a long metastable state before system's failure and finally a region of parameters for which the system is able to remain with a large fraction of nodes. Lastly, also note that the single layer setup constitutes an upper bound for the "OR" case, that is, the interdependent system cannot perform better for the "OR" rule than a single network with the same degree distribution.

\begin{figure}[htb]
\centering
\includegraphics[width=1.0\columnwidth]{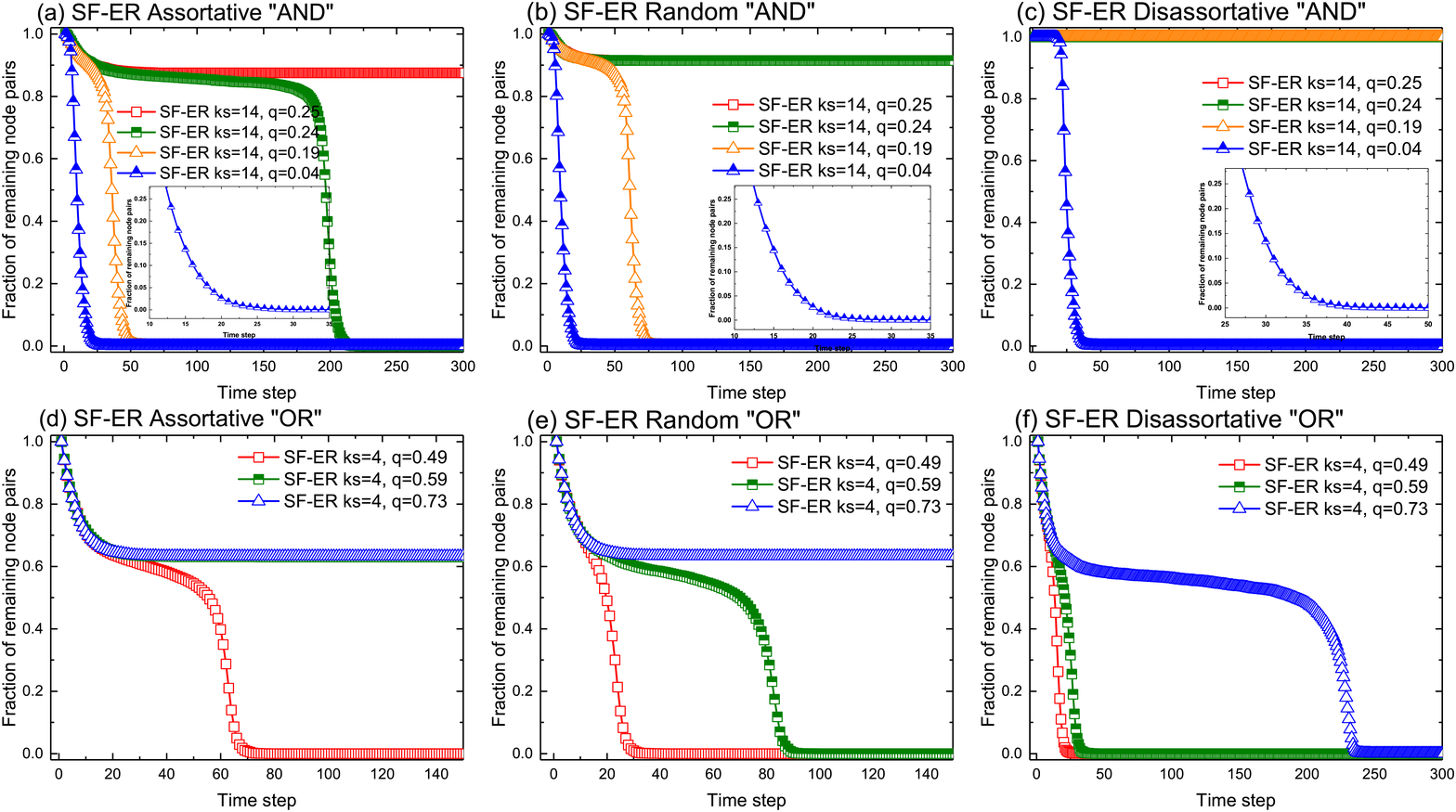}
\caption{Fraction of node pairs that remains in the whole network as a function of time for different model parameters and several configurations of the interdependent system as indicated. On the top three panels, the detachment rule for an interlayer pair to leave the network is the "AND" logic, while the "OR" logic was adopted for the three panels at the bottom. The probability of removing nodes one they meet the detachment rule has been set to $f=0.2$. The network parameters are as in Fig.\ref{fig:multitype_1}. From left to right, both in the top and the bottom rows, the degree correlations are assortative, random and disassortative.}
\label{fig:multitype_2}
\end{figure}

Next, we focus on the evolution of the fraction of remaining pairs on interdependent networks when the system is made with different network classes (i.e., $multitype=2$). Figure \ref{fig:multitype_2} shows results obtained for different interlayer degree correlations and for combinations consisting of a scale-free network and an ER graph. We have chosen this setup because it combines the more resilient with the more fragile topology, but we present the results obtained for the rest of combinations in the SI. As it can be clearly observed in the figure, the crashing behavior of the system is qualitatively consistent with the patterns previously discussed (see Fig.\ref{fig:multitype_1}). However, there are important quantitative differences as far as the impact of the removing rule or the role of the correlations concerns, see Fig.\ref{fig:multitype_2}. Focusing our attention in the top three panels, which correspond to the "AND" scenario, we observe that for the same $k_s$ in panel (a), the networks become much more fragile as $q$ decreases, reaching the point at which the whole system crashes with a small reduction in $q$. The same tendency can be observed also in panels (b) and (c). However, as clearly seen by comparing panels (a), (b) and (c), degree correlations do not always make the system more robust. Indeed, it depends on the kind of correlations: with respect to the random scenario, assortative correlations make the interdependent network less robust whereas disassortative correlations favor resilience. Interestingly enough, as seen in the bottom panels of Fig.\ref{fig:multitype_2}, the previous role of the correlations depends on the logic rule. Admittedly, the effect of correlations for the "OR" rule is in the opposite direction, that is, the more positively correlated the system is, the more robust it is. These results can be understood by noticing that the detachment condition is first fulfilled by lowly connected nodes regardless of whether the system operates under "AND" or "OR" rule. Thus, disassortative interdependent networks, in which lowly connected nodes of one layer are preferentially linked to nodes with a higher connectivity in the other layer, are harder to break for the "AND" rule because the pair of nodes has to meet the failure condition simultaneously, that is, the cascade is limited by the resilience of the highly connected nodes. This is not the case of the "OR", that is indeed ruled by the resilience of the low connected nodes as in this case, it is enough that one node in the pair fulfills the detachment condition.

\subsection*{$k_s$-$q_{th}$ curves and cascade sizes}

\begin{figure}[htb]
\centering
\includegraphics[width=1.0\columnwidth]{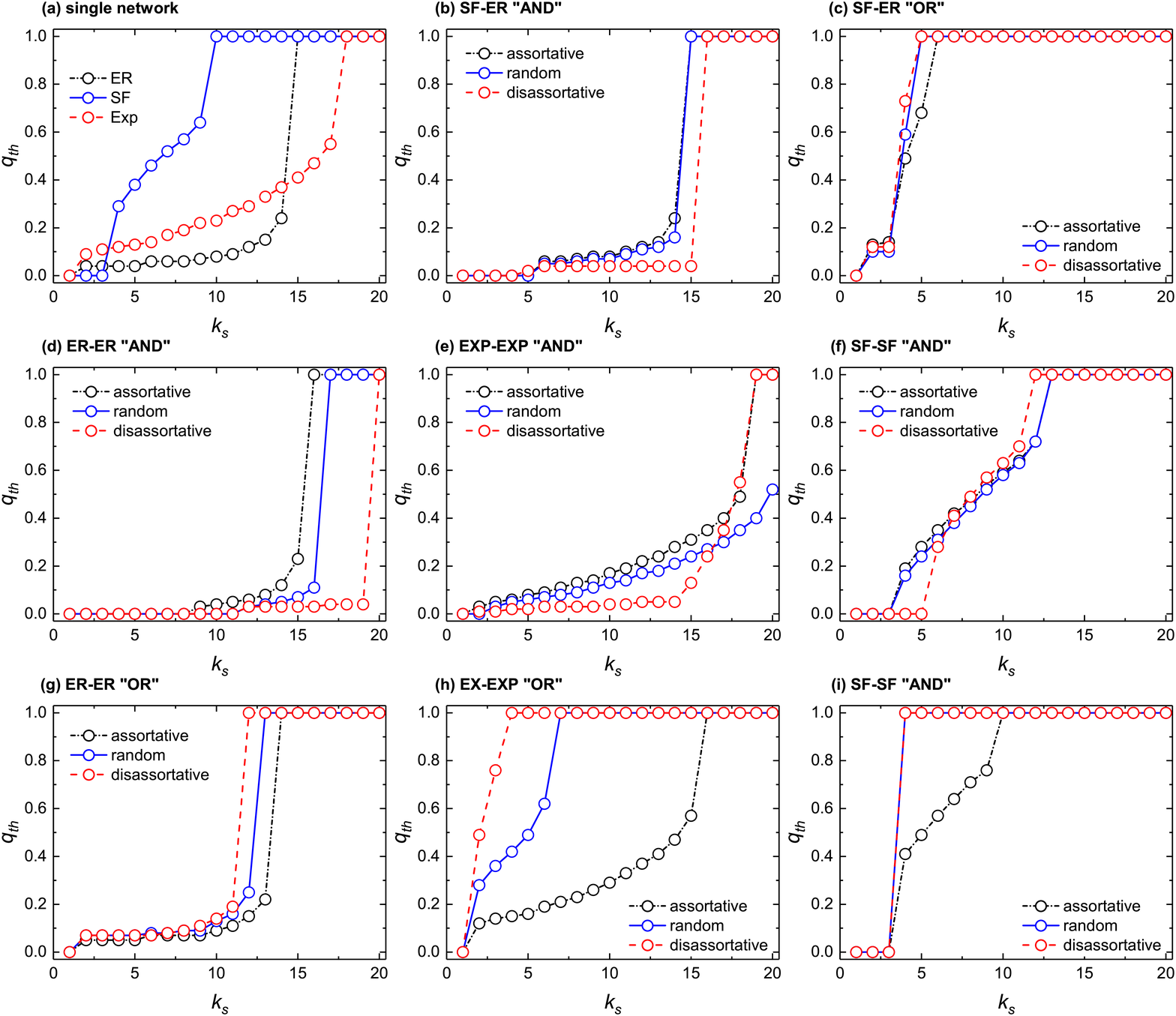}

\caption{Threshold $q_{th}$ of modified $KQ$ cascade model under different detachment rules, network setups and interlayer degree correlations. All networks have a size of $N = 10^4$ nodes and the rest of structural parameters are those used in Fig.\ref{fig:multitype_1} and Fig.\ref{fig:multitype_2}.}
\label{fig:k_q}
\end{figure}

The $KQ$-cascade model presents a critical line in the plane ($k_s,q$) that defines the critical thresholds. For instance, once $k_s$ is fixed, there is a critical value of $q_{th}$ such that the system will not collapse if $q> q_{th}$, whereas it breaks down for $q\le q_{th}$. Figure~\ref{fig:k_q} shows results for the $q_{th}$-$k_s$ curves obtained for single layer networks as well as for the interdependent networks studied here, including the different scenarios of degree correlations between the two layers and the two ("AND" and "OR") breaking rules. As it can be seen in this figure, the introduction of the second network coupled to the first one, i.e., the interdependent system, greatly changes the $k_s, q_{th}$ curves with respect to the single layer situation. Firstly, for the "AND" case, the whole system turns out to be more robust (not only with respect to the single layer scenario, but also when compared to the OR setting) and the $k_s$-$q_{th}$ curve shifts rightwards, meaning that there is a larger range of values of $k_s$ for which $q_{th}$ remains small. Secondly, we see that the influence of the degree correlations discussed previously is always present, but more noticeable for $multitype=1$ settings and the OR detachment rule [e.g., panels (g), (h) and (i)]. These results are consistent with those in Fig.\ref{fig:multitype_1} and Fig.\ref{fig:multitype_2}, which are already discussed. Moreover, we present the rest of cases in the SI.

\begin{figure}[htb]
\centering
\includegraphics[width=1.0\columnwidth]{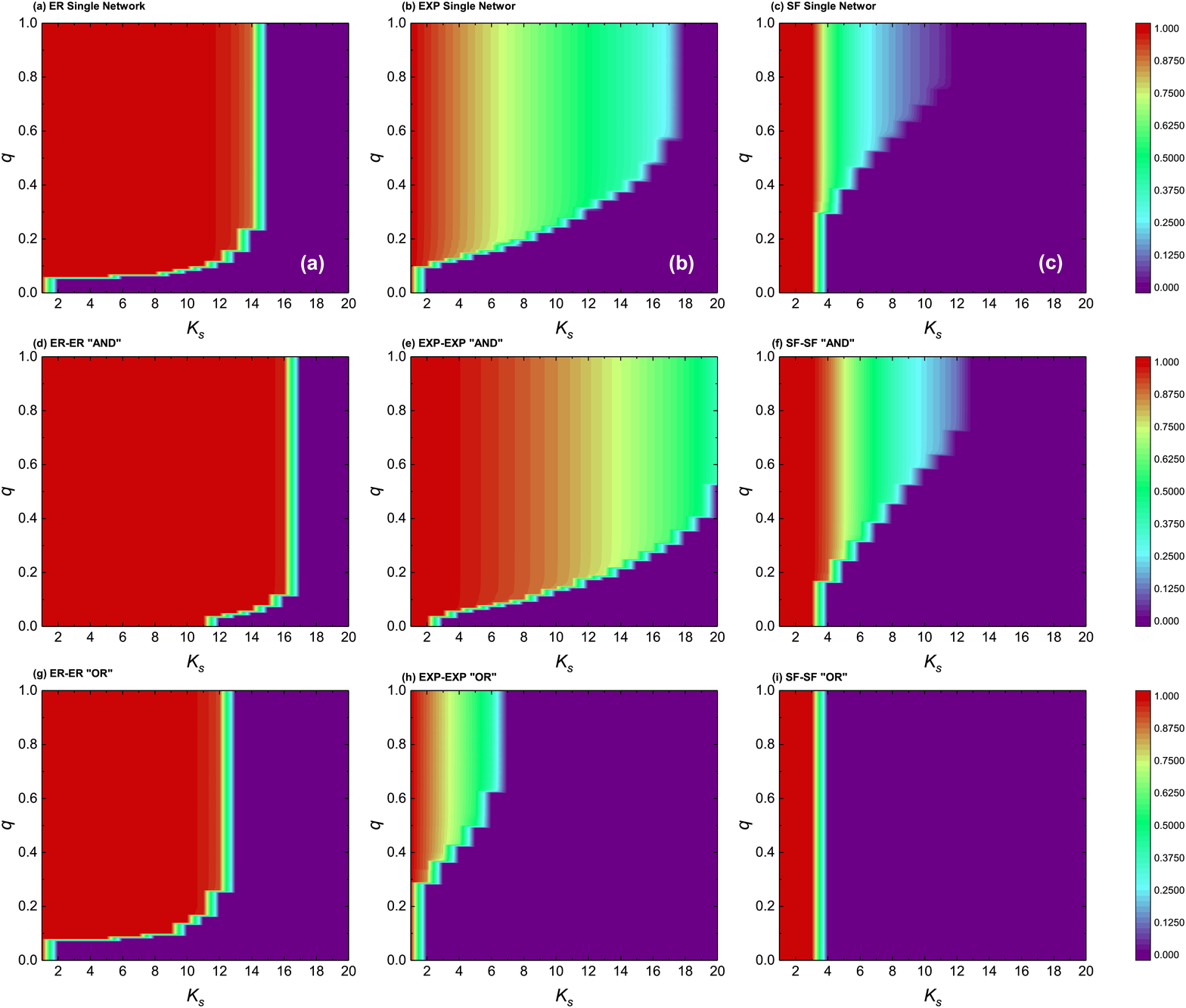}
\caption{Cascade size on different types of single and interdependent networks of size $N = 10^4$. The results correspond to different topologies (ER, exponential, scale-free architectures) and the same class of networks for each layer of the interdependent system. We also show results for the AND and OR detachment mechanism and interlayer degree correlations are neglected (i.e., the panels correspond to the random case). In all panels, the color denotes the fraction of remaining node pairs within the network after $500$ iterations of the dynamics. All other parameters are those used in Fig. \ref{fig:multitype_1}.}
\label{fig:cascade_size_multiType=2}
\end{figure}

\begin{figure}[htb]
\centering
\includegraphics[width=1.00\columnwidth]{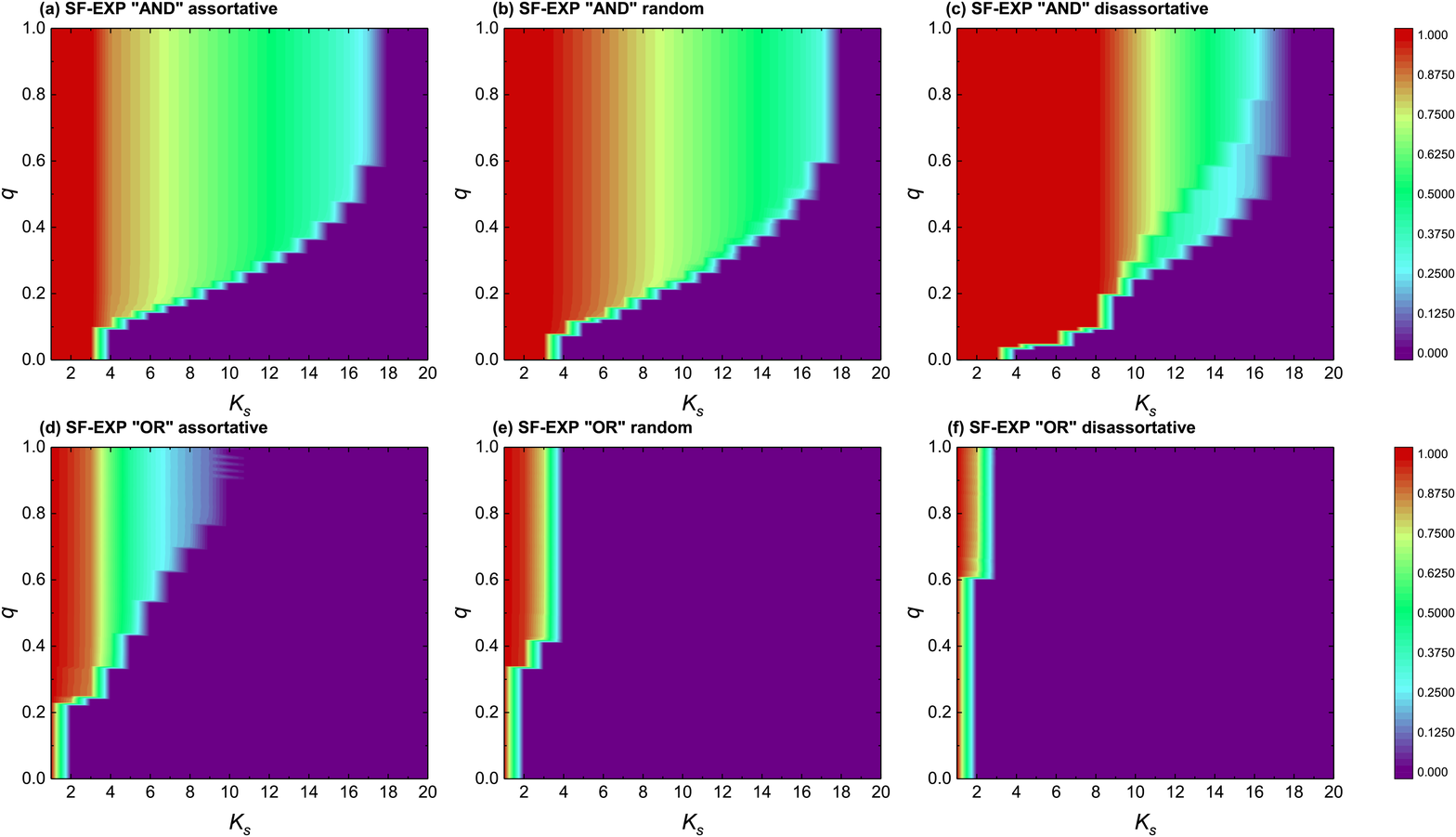}
\caption{Cascade size on different types of single and interdependent networks of size $N = 10^4$. The results correspond to different detachment rules and an SF-EXP combination of topologies for the interdependent system. We also show results for the AND and OR detachment mechanism and different interlayer degree correlations as indicated. In all panels, the color denotes the fraction of remaining node pairs within the network after $500$ iterations of the dynamics. All other parameters are those used in Fig. \ref{fig:multitype_1}.}
\label{fig:cascade_size_SF-SF}
\end{figure}

As mentioned above, the normalized size of a cascade is usually defined as the ratio between the number of nodes remaining at the final step and the initial number of nodes of the network, which is a parameter that characterizes the robustness of the network, namely, the higher the ratio is, the more resilient the system is. We next present our results regarding the evolution of this ratio for various types of networks and interlayer degree correlations studied so far as a function of both $k_s$ and $q$. In panels (a), (b) and (c) of Fig.~\ref{fig:cascade_size_multiType=2}, we first recover the results of Ref.\cite{ISI:000385610400047}, where the standard $KQ$-cascade model was studied on top of ER [panel (a)], exponential [panel (b)] and scale-free [panel (c)] networks. The comparison with results obtained for the interdependent networks shows, again, that the effect of the interdependency on the robustness of the system depends on the detachment scenario. While interdependent networks are more robust under the strict "AND" rule, they are not for the "OR" scenario. And this happens regardless of the network class, i.e., for ER, exponential and SF networks. Furthermore, we have also explored the same parameter space of Fig.~\ref{fig:cascade_size_multiType=2}, but for correlated interdependent networks with different classes of networks in each layer, see Fig.\ref{fig:cascade_size_SF-SF}. Here we show results for  the $SF-EXP$ network setup for the sake of simplicity (results for the other two combinations are provided in the SI). Again, we see that the interlayer degree correlations play a significant role in the system's robustness and dynamics. Likewise, for this setup, the differences between the AND and the OR scenarios are enhanced [for instance, compare the difference between panels (c) and (f) in Fig.\ref{fig:cascade_size_SF-SF} with that of panels (f) and (i) of Fig.~\ref{fig:cascade_size_multiType=2}].

\section*{Conclusions}
\label{sec:4}

In summary, in this paper we have extended the standard $KQ$ cascade model to the more realistic scenario of interdependent networks. It is worth mentioning that we have found that several dynamical outcomes are still observed for these systems: sudden system's breakdown, metastable states and highly robust regimes. Having at least two layers made it possible to also introduce two different scenarios for the detachment of the nodes. The first one, more restrictive "AND" rule, imposes that a node fails if and only if it and its corresponding partner simultaneously fulfill the breakdown ($KQ$) condition. Conversely, the "OR" rule only requires that any of the two nodes fulfills the $KQ$-condition. Moreover, we also studied different interlayer correlations and topological configurations of the networks that make up the whole system. Our results show that both correlations and the detachment rules have a great impact on the resilience of the interdependent system. In general, the "AND" rule leads to more robust scenarios, whereas the "OR" case makes the system more fragile. Interestingly enough, the effect of coupling correlations on the robustness of the system depends on the detachment rule: for the "AND" case, the more disassortative the system is, the more resilient it becomes. On the contrary, the more positively correlated the two layers of the interdependent system are, the more robust the system is for the "OR" scenario. This makes the dynamics on top of this two-layer systems highly nontrivial. Taking together, our results show that within the $KQ$-cascade model, the fragility or robustness of networked systems is related not only to their topological structure, but also to the rules governing the cascading dynamics and the mixing interlayer patterns. Thus, our results might provide deeper insights into the properties of cascading process in real-world systems.

\bibliography{sample}

\section*{Acknowledgements}
This project is financially supported by the National Natural Science Foundation of China (NSFC) (under Grants 61773286 and 61374169), and C.X. also acknowledges the funding support of China Scholarship Council (under Grant 201808120001). Y.M. acknowledges partial support from the Government of Arag\'on, Spain through grant E36-17R (FENOL), by MINECO and FEDER funds (FIS2017-87519-P) and by Intesa Sanpaolo Innovation Center. Additionally, we thank a lot to Ms. Yaning Zhao for her huge efforts to the revised edition.

\section*{Author contributions statement}
G.X. and C.X. conceived and designed the study, J.L. and C.X. conducted the simulations. All authors analyzed the results, wrote and reviewed the manuscript.

\section*{Additional information}
\textbf{Competing interests}: The authors declare no competing interests.

\end{document}